\documentclass[a4paper,11pt]{article}
\usepackage{pos}
\usepackage{braket}
\usepackage{CJKutf8}

\title{Extracting the distribution amplitude of pseudoscalar mesons using the HOPE method}

\author[a]{S.-P. Alex Chang}
\emailAdd{s44930e0@gmail.com}
\affiliation[a]{Institute of Physics, National Yang Ming Chiao Tung University, \\
1001 Ta-Hsueh Road, Hsinchu 30010, Taiwan}

\author[b]{William Detmold}
\emailAdd{wdetmold@mit.edu}
\affiliation[b]{Center for Theoretical Physics, Massachusetts Institute
  of Technology, \\
Cambridge, MA 02139, USA}

\author[c]{Anthony V. Grebe}
\emailAdd{agrebe@mit.edu}
\affiliation[c]{Fermi National Accelerator Laboratory, \\
Batavia, IL 60502, USA}

\author[d]{Issaku Kanamori}
\emailAdd{kanamori-i@riken.jp}
\affiliation[d]{RIKEN Center for Computational Science, \\
Kobe 650-0047, Japan}

\author[a,e]{C.-J. David Lin}
\emailAdd{dlin@nycu.edu.tw}

\affiliation[e]{Centre for High Energy Physics, Chung-Yuan Christian University, \\
Chung-Li, 32032, Taiwan}

\author*[b,f]{Robert J. Perry}
\emailAdd{perryrobertjames@gmail.com}
\affiliation[f]{Departament de F\'isica Qu\`antica i Astrof\'isica and Institut de Ci\`encies del Cosmos, Universitat de Barcelona, \\
Mart\'i Franqu\`es 1, E08028, Spain}

\author[g]{Yong Zhao}
\emailAdd{yong.zhao@anl.gov}
\affiliation[g]{Physics Division, Argonne National Laboratory, \\
Lemont, IL 60439, USA}

\abstract{The pseudoscalar meson light-cone distribution amplitudes (LCDAs) are essential non-perturbative inputs for a range of high-energy exclusive processes in quantum chromodynamics. In this proceedings, progress towards a determination of the low Mellin moments of the pion and kaon LCDAs by the HOPE Collaboration is reported.}

\FullConference{The 41st International Symposium on Lattice Field Theory (LATTICE2024)\\
 28 July - 3 August 2024\\
Liverpool, UK\\}

\begin{document}
\maketitle

\section{Introduction}
The light-cone distribution amplitude (LCDA) is a non-perturbative object of interest for a range of high-energy exclusive processes in quantum chromodynamics (QCD)~\cite{Lepage:1980fj}. It can be defined in terms of the matrix element,
\begin{equation}
\label{eq:pion_DA_def}
\braket{ 0 | \overline{\psi}(z) \gamma^{\mu} \gamma_{5} W[z, -z] \psi(-z) |
M^{+} ({\vec{p}}) } =
 i  p^{\mu} f_{M} \int_{-1}^{1} d \xi \mbox{ }
 {\mathrm{e}}^{-i \xi p\cdot z }\phi_{M}(\xi, \mu ),
\end{equation}
where ${\mathcal{W}}[-z,z]$ is a light-like ($z^{2} = 0$) Wilson line and $\phi_{M}(\xi, \mu )$ is the LCDA for the meson $M$.  In the above equation, $f_{M}$, ${\vec{p}}$, and $p_{\mu}$ are the decay constant, three-momentum, and the four-momentum of the meson, respectively.

Despite the clear phenomenological value in computing this quantity from first principles, the presence of a light-like separation in the definition of the relevant operator prohibits a direct calculation using Euclidean formulations of lattice QCD (LQCD). Historically, the solution has been to employ an operator product expansion (OPE) to relate this operator to a sum of local operators, which can be calculated using LQCD~\cite{Kronfeld:1984zv,Martinelli:1987si}. The meson-to-vacuum matrix elements of these local operators can be shown to be related to the Mellin moments of the LCDA, 
\begin{equation}
\braket{ 0 | \overline{\psi}\gamma^{\{\mu_1}\gamma_5(i\overset{\leftrightarrow}{D}\vphantom{D}^{\,\mu_2})\dots(i\overset{\leftrightarrow}{D}\vphantom{D}^{\,\mu_n\}})\psi |
M^{+} ({\vec{p}}) } = f_M \braket{\xi^n}_M(\mu) [p^{\mu_1} p^{\mu_2}\dots p^{\mu_n}-\text{tr}],
\end{equation}
where
\begin{equation}
\braket{\xi^n}_M(\mu)=\int_{-1}^{1} d\xi\, \xi^n \phi_M(\xi,\mu)
\end{equation}
are the Mellin moments. This approach has been limited to the lowest few moments because the higher spin lattice operators mix with lower dimensional operators and the mixing coefficients contain power divergences. More recently, other approaches have been proposed~\cite{Aglietti:1998ur,Liu:1999ak,Detmold:2005gg,Braun:2007wv,Davoudi:2012ya,Ji:2013dva,Chambers:2017dov,Radyushkin:2017cyf,Ma:2017pxb,Shindler:2023xpd,Gao:2023lny}. 

\section{The HOPE method}
The heavy-quark operator product expansion (HOPE) is an alternative approach~\cite{Detmold:2005gg,Detmold:2021uru} to extracting information about the Mellin moments of the LCDA. It is based on studying the hadronic matrix element
\begin{align}
  R_M^{\mu\nu}(t_-, \vec{p}, \vec{q})
 &= \int d^3 \vec{z} \,e^{i\vec{q}\cdot \vec{z}} \braket{ 0 |T\{ J_A^\mu(t_-/2,\vec{z}/2) J_A^\nu(-t_-/2,-\vec{z}/2)\} | M(\vec{p}) },
\label{eq:ratio}
\end{align}
where 
\begin{equation}
J_{A}^{\mu}=\overline{\Psi} \gamma^{\mu}\gamma^{5}\psi + \overline{\psi} \gamma^{\mu}\gamma^{5} \Psi    
\end{equation}
is an axial-vector current involving a light or strange quark $\psi$ and a fictitious valence heavy quark $\Psi$ with mass $m_{\Psi}$.
The momentum-space hadronic matrix element is given by
\begin{equation}
V_M^{\mu\nu}(p,q) = \int dt_- e^{iq_4 t_-} R_M^{\mu\nu}(t_-, \vec{p}, \vec{q}).
\end{equation}
The matrix element of this two-current operator can be written in terms of local operators via an application of the heavy-quark operator product expansion. To one-loop it can be written as~\cite{Detmold:2021uru}
\begin{equation}
\label{eq:Gegen_OPE_had_amp}
 V_M^{\mu\nu} (q,p) = - \frac{2 i  f_{\pi}\epsilon^{\mu\nu\rho\sigma}
 q_{\rho} p_{\sigma}}{\tilde{Q}^{2}}  
 \sum_{n=0,{\mathrm{even}}}^{\infty} {\mathcal{F}}_{n} (\tilde{Q}^{2},
 \mu, \tilde{\omega}, m_\Psi) \phi_{M,n} (\mu)  + 
  \text{higher-twist terms}\, ,
\end{equation}
where $\phi_{M,n} (\mu)$ are the Gegenbauer moments, $\mu$ is the renormalization scale, and ${\mathcal{F}}_{n}$  are coefficients that can be computed in QCD perturbation theory and can be expressed as functions of the kinematic variables,
\begin{equation}
\label{eq:kin_var_def}
 \tilde{Q}^{2} = Q^{2} + m_{\Psi}^{2} \, , ~~~~~
 \tilde{\omega} = \frac{2 p\cdot q}{\tilde{Q}^{2}} \, ,
\end{equation}
with $Q^{2} = -q^{2}$. To zeroth order in the coupling, the above expression can be equivalently written as
\begin{equation}
V_M^{\mu\nu} (q,p) = - \frac{2 i  f_{\pi}\epsilon^{\mu\nu\rho\sigma}
 q_{\rho} p_{\sigma}}{\tilde{Q}^{2}}  
 \sum_{n=0}^{\infty} \bigg(\frac{\tilde{\omega}}{2}\bigg)^n \braket{\xi^n}_M + 
  \text{higher-twist terms}\, ,
\end{equation}
where the Gegenbauer and Mellin moments of the LCDA are related via
\begin{equation}
\begin{split}
\label{eq:gegen_to_mellin}
\phi_{0,M}=\braket{\xi^0}_M=1, 
~~~~
\phi_{1,M}=\frac{5}{3}\braket{\xi^1}_M, 
~~~~
\phi_{2,M}=\frac{7}{12}\left(5\braket{\xi^2}_M-\braket{\xi^0}_M\right),
\\
\phi_{3,M}=\frac{3}{4}\left(7\braket{\xi^3}-3\braket{\xi^1}\right),
~~~~
\phi_{4,M}=\frac{11}{24}\left(21\braket{\xi^4}_M - 14\braket{\xi^2}_M + \braket{\xi^0}_M\right),
~~~~
\dots
\end{split}
\end{equation}

\section{Numerical Strategy}
In practice, the HOPE strategy involves computing correlation functions using LQCD which provide access to the hadronic matrix element defined by Eq.~\eqref{eq:ratio}. By fitting the numerical data to the HOPE formula given by Eq.~\eqref{eq:Gegen_OPE_had_amp}, information about the Gegenbauer moments of the meson LCDA can be determined. From the relations given in Eq.~\eqref{eq:gegen_to_mellin}, this information can be converted into information about the Mellin moments. The hadronic matrix element can be determined from the correlation functions,
\begin{align}
C_{ij}(t,\vec{p})&=\int d^3x e^{i\vec{p}\cdot \vec{x}} \braket{0|\mathcal{O}_i(t,\vec{x})\mathcal{O}_j^\dagger(0,\vec{0})|0}\,,
\\
C_{j}^{\mu\nu}(t_e,t_m,\vec{p}_e,\vec{p}_m)&=\int d^3x_e d^3x_m e^{i\vec{p}_e\cdot \vec{x}_e} e^{i\vec{p}_m\cdot \vec{x}_m} \braket{0|J_A^\mu(t_e,\vec{x}_e)J_A^\nu(t_m,\vec{x}_m)\mathcal{O}_j^\dagger(0,\vec{0})|0}\,,
\end{align}
where $\mathcal{O}_i(t,\vec{x})$ is an interpolating operator for the meson. In the studies covered in this proceedings, two interpolating operators, $\mathcal{O}_1=\overline{\psi}\gamma_5 \psi $ and $\mathcal{O}_2=\overline{\psi}\gamma_4\gamma_5 \psi$, are employed.
The optimal linear combination which maximizes the overlap of the operator onto the ground state is constructed by solving the generalized eigenvalue problem~\cite{Michael:1982gb,Luscher:1990ck},
\begin{equation}
\sum_{j}{C}_{ij}(t_\text{ref},\vec{p}){v}_{n,j}(t_\text{ref},t_0)=\lambda_n(t_\text{ref},t_0)\sum_{j}{C}_{ij}(t_0,\vec{p}){v}_{n,j}(t_\text{ref},t_0).
\end{equation}
The eigenvector ${v}_{n,j}(t_\text{ref},t_0)$ defines the optimized interpolating operator for the $n$th state. Thus the optimized interpolating operator for the ground state is
\begin{equation}
\mathcal{O}_M(t,\vec{x}) = \sum_{j} {v}_{0,j}(t_\text{ref},t_0) \mathcal{O}_j(t,\vec{x}).
\end{equation}

This optimized interpolating operator is used self-consistently in the definitions of the two- and three-point correlation functions. In terms of this optimized interpolating operator for the pion, the two-point correlation function $C_{MM}(t,\vec{p})$ admits the spectral decomposition
\begin{equation}
C_{MM}(t,\vec{p})=\sum_n \frac{|Z_{n,M}(\vec{p})|^2}{2E_n} e^{-E_n(\vec{p}) t}  
\end{equation}
where $Z_{n,M}(\vec{p}) = \braket{0|\mathcal{O}_M(0,\vec{0})|n,\vec{p}}$.

The three-point correlation function $C_M$ can be written as
\begin{equation}
C_{M}^{\mu\nu}(t_e,t_m,\vec{p}_e,\vec{p}_m) = R_M^{\mu\nu} (t_-,\vec{p},\vec{q}) \frac{Z_{0,M}^*(\vec{p})}{2E_0(\vec{p})}  e^{-E_0(\vec{p})t_+/2} + \text{excited states},
\end{equation}
where $t_+=t_e+t_m$, $t_-=t_e-t_m$, $\vec{p}=\vec{p}_e+\vec{p}_m$ and $\vec{q}=(\vec{p}_e-\vec{p}_m)/2$. By constructing the ratio
\begin{equation}
\mathcal{R}_M^{\mu\nu}(t_-,t_+,\vec{p},\vec{q}) = \frac{C_{M}(t_e,t_m,\vec{p}_e,\vec{p}_m)}{\frac{Z_{0,M}(\vec{p})}{2E_0(\vec{p})}  e^{-E_0(\vec{p})t_+/2}},
\end{equation}
and studying the large $t_+$ limit, it is possible to show that
\begin{equation}
\lim_{t_+\to\infty}\mathcal{R}_M^{\mu\nu}(t_-,t_+,\vec{p},\vec{q}) = R_M^{\mu\nu}(t_-,\vec{p},\vec{q}).
\end{equation}
which may be fit to the HOPE formula to obtain information about the low Mellin moments of the meson LCDA.

\section{Results}
\subsection{Determination of fourth Mellin moment of pion in the quenched approximation}
In Ref.~\cite{Detmold:2021qln}, the second moment of the pion LCDA was determined using the HOPE method. However, the pion momentum was insufficient to study higher moments of the pion LCDA. The following analysis builds upon that work by studying the hadronic matrix element $R^{\mu\nu}(t,\vec{p},\vec{q})$ at larger pion three-momenta. In this study, the pion state is taken to have $|\vec{p}|/(2\pi/L)=2$ units of momentum.
The hadronic matrix element is computed on four ensembles generated in the quenched approximation for a range of heavy quark masses. The simulation parameters for these ensembles have previously been reported in Ref.~\cite{Detmold:2021qln}.

\begin{figure}
    \centering
    \includegraphics[width=0.45\linewidth]{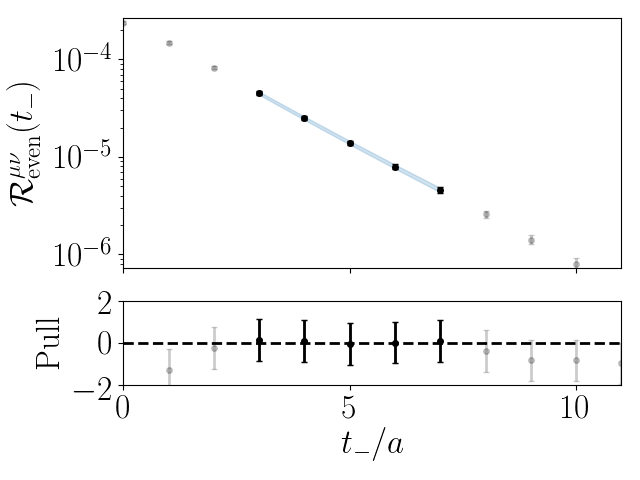}
    \includegraphics[width=0.45\linewidth]{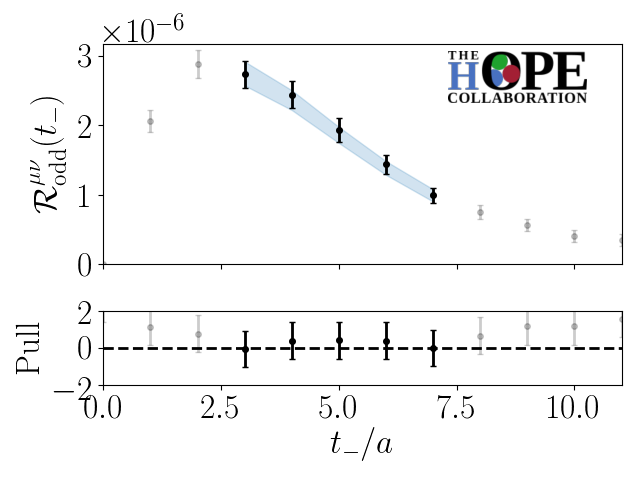}
    \caption{Fitting one-loop HOPE expression to the $t_-$-even and $t_-$-odd components of the hadronic matrix element computed using LQCD.}
    \label{fig:xi4_hadronic_matrix_element}
\end{figure}

\begin{figure}
    \centering
    \includegraphics[width=0.45\linewidth]{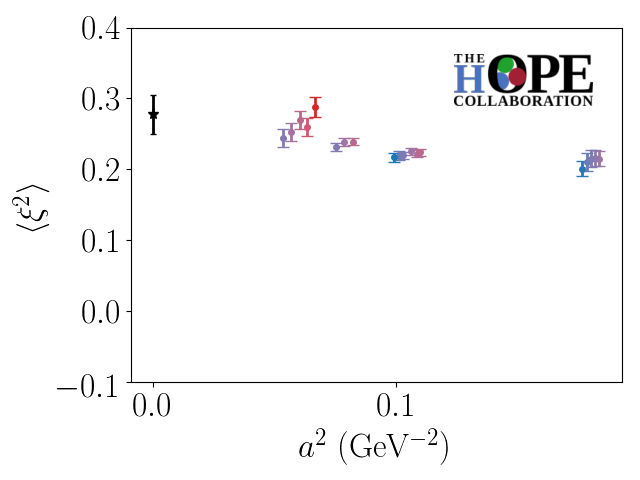}
    \includegraphics[width=0.45\linewidth]{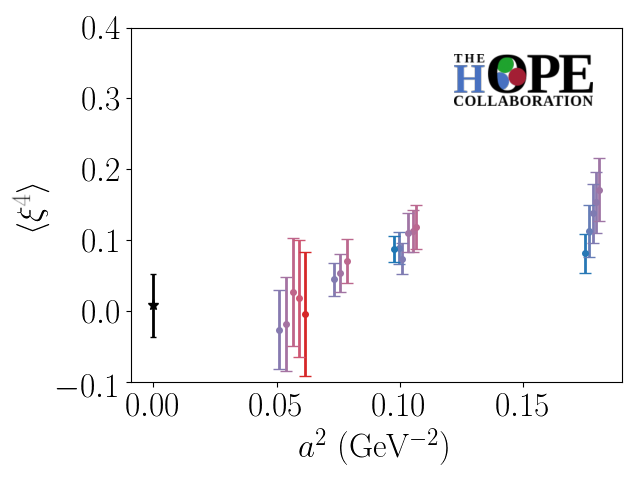}
    \caption{Performing continuum, twist-2 extrapolation of fitted $\braket{\xi^2}$.}
    \label{fig:xi4_continuum_twist_2_extrapolation}
\end{figure}

In this study the ratio $\mathcal{R}_\pi^{\mu\nu}(t_-,t_+,\vec{p},\vec{q})$ is fit to the one-loop HOPE formula evaluated at a renormalization scale of $\mu=2$~GeV. An example of this fit is shown in Fig.~\ref{fig:xi4_hadronic_matrix_element}. Following this, data is extrapolated to the continuum, twist-two limit using the extrapolation formula,
\begin{equation}
\label{eq:continuum_twist_two}
X(a,m_\Psi) = X_0 + \frac{A}{m_\Psi} + B a^2 + C a^2 m_\Psi + D a^2m_\Psi^2,
\end{equation}
where $X\in\{f_\pi, \braket{\xi^2}, \braket{\xi^4}\}$.
The form of this extrapolation is inspired by power-counting arguments for the leading higher-twist and lattice artifacts. In order to estimate the systematic uncertainty stemming from the use of this parameterization, the model averaging prescription proposed in Ref.~\cite{Jay:2020jkz} is employed. The resulting continuum, twist-two extrapolations of the second and fourth Mellin moments are presented in Fig.~\ref{fig:xi4_continuum_twist_2_extrapolation}. 

\subsection{Progress in determination of second Mellin moment of pion}
Previous work~\cite{Detmold:2021qln} focused on a proof-of-principle determination of the second moment of the pion LCDA. Due to the exploratory nature of that work, the quenched approximation was employed. As a result, the determined second Mellin moment of the LCDA suffers from a uncontrolled systematic error. Therefore, the HOPE Collaboration is currently repeating this calculation using dynamical ensembles generated by the CLS Collaboration~\cite{RQCD:2022xux}. Details of the dynamical ensembles employed in this work are given in Table~\ref{tab:CLS_ensembles}. Using this set of ensembles, it will be possible to perform an extrapolation to the physical point in addition to the continuum, twist-two extrapolation performed in previous HOPE studies. 

\begin{table}
\centering
\renewcommand{\arraystretch}{1.2}
\begin{tabular}{ c c c c c c c c} 
\hline \hline 
Name & $(L/a)^3 \times T/a$ & $\beta$ & $a$ (fm) & $\kappa_\text{light}$ & $\kappa_\text{strange}$ & $m_\pi$ (MeV) & $m_K$ (MeV)
\\ \hline
B451 & $32^3 \times 64$ & 3.46 & 0.075 & 0.136981 & 0.136409 & 422 & 577 
\\
N305 & $48^3 \times 128$ & 3.7 & 0.049 & 0.137025 & 0.136676 & 428 & 584
\\
H107 & $32^3 \times 96$ & 3.4 & 0.085 & 0.136946 & 0.136203 & 368 & 550
\\
B452 & $32^3 \times 64$ & 3.46 & 0.075 & 0.137046 & 0.136378 & 352 & 548
\\
N204 & $48^3 \times 128$ & 3.55 & 0.064 & 0.137112 & 0.136575 & 353 & 549
\\
N304 & $48^3 \times 128$ & 3.7 & 0.049 & 0.137079 & 0.136665 &353 & 558
\\
N450 & $48^3 \times 128$ & 3.46 & 0.075 & 0.137099 & 0.136353 & 287 & 528
\\
\hline \hline 
\end{tabular}
\caption{Lattice details for dynamical ensembles employed in calculation of second Mellin moment of the pion and low Mellin moments of the kaon. Further details about these ensembles can be found in Ref.~\cite{RQCD:2022xux}}
\label{tab:CLS_ensembles}
\end{table}

\begin{figure}
    \centering
    \includegraphics[width=0.45\linewidth]{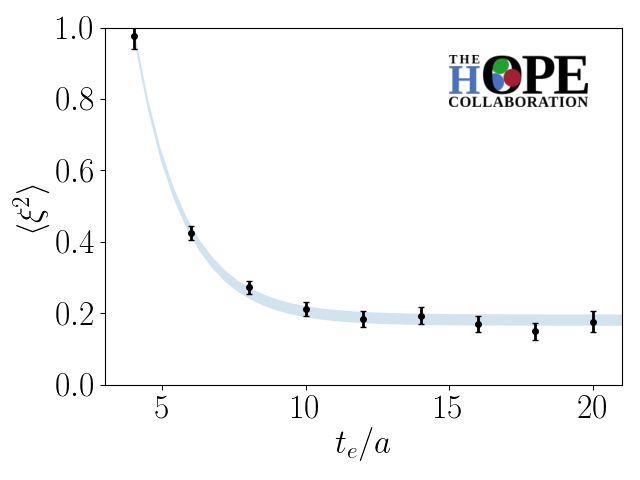}        \includegraphics[width=0.45\linewidth]{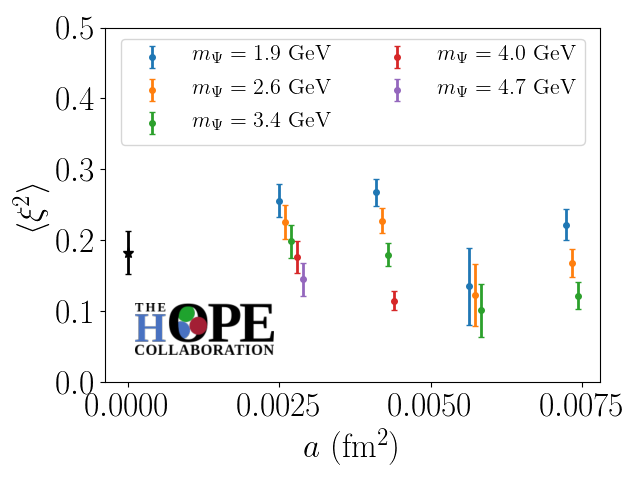}
    \caption{(Left) Studying excited state contamination in the extracted second Mellin moment of the pion LCDA. (Right) Continuum, twist-two extrapolation of the second moment of the pion LCDA using ensembles with a pion mass of approximately $m_\pi\approx350$~MeV.}
    \label{fig:xi2_dynamical}
\end{figure}

Except in several small ways, this analysis mirrors the analysis of the fourth moment and the previous quenched analysis of the second moment~\cite{Detmold:2021qln}. A preliminary study of excited state contamination in the extracted second Mellin moment is presented in Fig.~\ref{fig:xi2_dynamical}. After extrapolating excited state contamination, the resulting values of the second moment on the ensembles considered here is shown in Fig.~\ref{fig:xi2_dynamical}, along with the preliminary continuum, twist-two extrapolation. This extrapolation employs the same \textit{ansatz} proposed in Eq.~\eqref{eq:continuum_twist_two}. It is important to note that this analysis only considers ensembles with approximately equivalent pion masses.

\subsection{Progress in determination of low moments of kaon LCDA}
\begin{figure}
    \centering
    \includegraphics[width=0.45\linewidth]{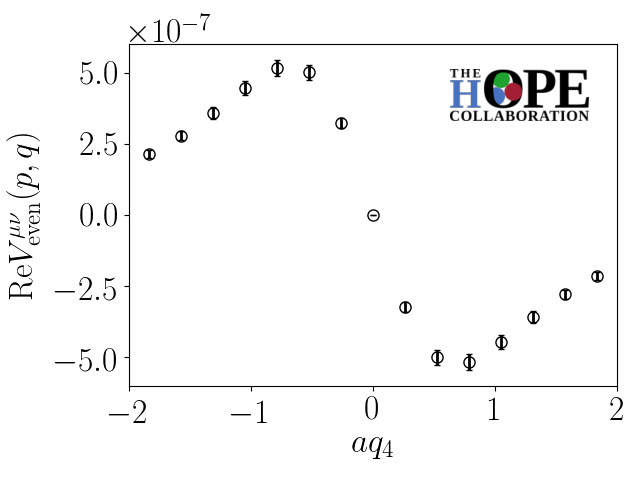}
    \includegraphics[width=0.45\linewidth]{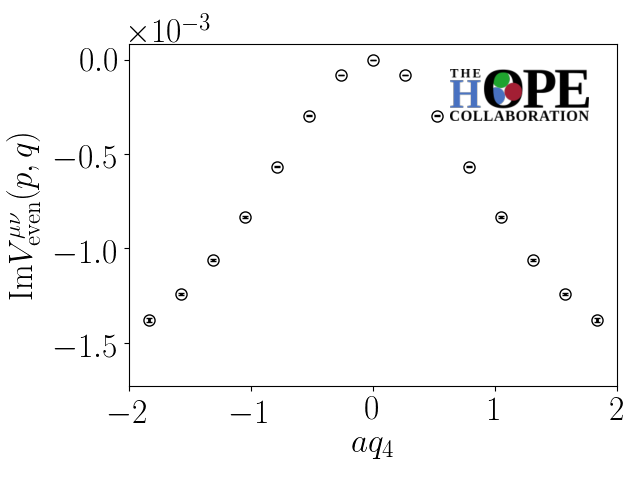}

    \includegraphics[width=0.45\linewidth]{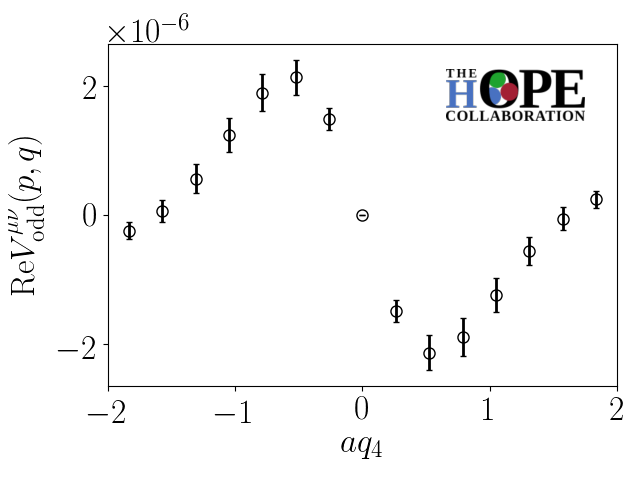}
    \includegraphics[width=0.45\linewidth]{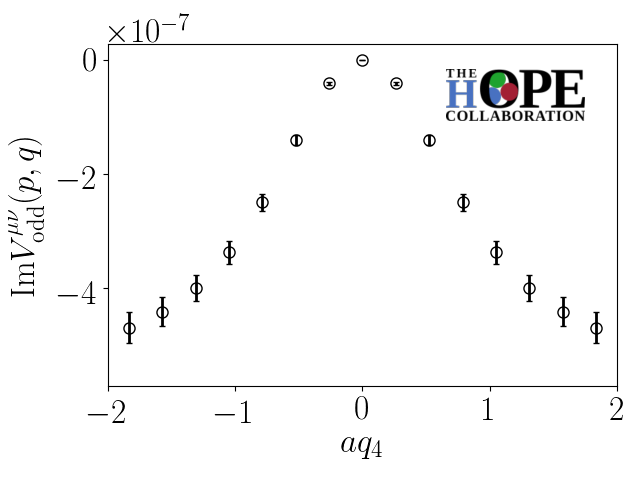}
    \caption{Real and imaginary components of the $q$-even and $q$-odd amplitudes which are sensitive to the even and odd Mellin moments of the Kaon LCDA.}
    \label{fig:kaon_LCDA}
\end{figure}

Exploratory work has commenced on the determination of the low moments of the kaon LCDA using the HOPE method. Due to the presence of a strange quark, both the odd and even Mellin moments are non-zero. In order to assess the sensitivity of the hadronic matrix element to the presence of the low Mellin moments it is useful to note that in momentum space, the linear combinations 
\begin{align}
V_{K,\text{even}}^{\mu\nu}(p,q)&=\frac{1}{2}\bigg[V_K^{\mu\nu}(p,q)-V_K^{\mu\nu}(p,-q)\bigg]\,,
\\
V_{K,\text{odd}}^{\mu\nu}(p,q)&=\frac{1}{2}\bigg[V_K^{\mu\nu}(p,q)+V_K^{\mu\nu}(p,-q)\bigg]\,,
\end{align}
are sensitive to the even and odd Mellin moments, respectively. That is, at tree-level,
\begin{align}
V_{K,\text{even}}^{\mu\nu}(p,q) = - \frac{2 i  f_{\pi}\epsilon^{\mu\nu\rho\sigma}
 q_{\rho} p_{\sigma}}{\tilde{Q}^{2}}  
 \sum_{n=0,\text{even}}^{\infty} \bigg(\frac{\tilde{\omega}}{2}\bigg)^n \braket{\xi^n}_M + 
  \text{higher-twist terms}\, ,
  \\
 V_{K,\text{odd}}^{\mu\nu}(p,q) = - \frac{2 i  f_{\pi}\epsilon^{\mu\nu\rho\sigma}
 q_{\rho} p_{\sigma}}{\tilde{Q}^{2}}  
 \sum_{n=0,\text{odd}}^{\infty} \bigg(\frac{\tilde{\omega}}{2}\bigg)^n \braket{\xi^n}_M + 
  \text{higher-twist terms}\, , 
\end{align}
In addition to this linear combination, it is possible to further separate the hadronic matrix element into real and imaginary parts, as previously employed in studies of the pion LCDA~\cite{Detmold:2021qln}. The combination of these two approaches make it possible to isolate $\braket{\xi^1}_K$ and $\braket{\xi^2}_K$ in $\text{Im}V_\text{odd}^{\mu\nu}(p,q)$ and  $\text{Re}V_\text{even}^{\mu\nu}(p,q)$, respectively. These amplitudes have been calculated using the ensembles listed in Table~\ref{tab:CLS_ensembles} and are shown in Fig.~\ref{fig:kaon_LCDA}. The clear non-zero signal for each of these amplitudes is encouraging for a determination of these two Mellin moments using the HOPE method.

\section{Conclusions}
The LCDA is the non-perturbative object of interest for a range of high-energy exclusive processes in QCD. In this proceedings, recent progress by the HOPE calculation in computing the low Mellin moments of the pion and the kaon LCDA are presented. 

\section*{Acknowledgements}

WD and RJP are supported in part by the U.S. Department of Energy, Office of
Science under grant Contract Number DE-SC0011090 and by the SciDAC5 award DE-SC0023116.
WD is additionally supported by the National Science Foundation under Cooperative Agreement PHY-2019786 (The NSF AI Institute for Artificial Intelligence and Fundamental Interactions, http://iaifi.org/). 
RJP has been supported by the projects CEX2019-000918-M (Unidad de Excelencia “Maria de Maeztu”), PID2020-118758GB-I00, financed by MICIU/AEI/10.13039/501100011033/ and FEDER, UE, as well as by the EU STRONG-2020 project, under the program H2020-INFRAIA-2018-1 Grant Agreement No. 824093.
AC and CJDL are supported by 112-2112-M-A49-021-MY3 and 113-2123-M-A49-001-SVP.
The authors thank ASRock Rack Inc.~for their support of the construction of an Intel Knights Landing cluster at National Yang Ming Chiao Tung University, where the numerical calculations were performed.
Help from Balint Joo in tuning Chroma is acknowledged.
The authors thankfully acknowledge the computer resources at MareNostrum and the technical support provided by BSC (RES-FI-2023-1-0030).
The authors gratefully acknowledge the support of ASGC (Academia Sinica Grid Computing Center, AS-CFII-114-A11, NSTC(NSTC 113-2740-M-001-007) for provision of computational resources.
This work utilized Supercomputer Fugaku in RIKEN Center for Computational Science
through the HPIC project hp220312 and hp230466.
We used Bridge++ code set~\cite{Bridge, Ueda:2014rya, Akahoshi:2021gvk}.
We thank Enno Scholz, Gunnar Bali, and the CLS Collaboration for generously sharing the ensembles used in the dynamical calculation in this work.
This document was prepared using the resources of the Fermi National Accelerator Laboratory (Fermilab), a U.S. Department of Energy, Office of Science, Office of High Energy Physics HEP User Facility. Fermilab is managed by Fermi Forward Discovery Group, LLC, acting under Contract No. 89243024CSC000002.
This material is based upon work supported by the U.S. Department of Energy, Office of Science, Office of Nuclear Physics through Contract No.~DE-AC02-06CH11357.

\bibliographystyle{JHEP}
\bibliography{zbibliography}

\end{document}